# Spin-Hall-Like Magnon Transport in a Synthetic Antiferromagnetic Skyrmion Lattice

Xingen Zheng[1,2*], Xuejuan Liu[3*], Zhenyu Wang[4], Jiyong Kang[1,2], Zhixiong Li[5+], and Hao Wu[1,2+]

[1]Dongguan Institute of Materials Science and Technology, Chinese Academy of Sciences, Guangdong 523808, China

[2]Songshan Lake Materials Laboratory, 523808, Dongguan, Guangdong, China

[3]Shenzhen Key Laboratory of Ultraintense Laser and Advanced Material Technology, Center for Intense Laser Application Technology, and College of Engineering Physics, Shenzhen Technology University, Shenzhen 518118, China

[4]School of Physics and Electronics, Hunan University, Changsha 410082, China

[5]School of Physics, Central South University, Changsha 410083, China

[*]These authors contributed equally to this work.

[+]Author to whom any correspondence should be addressed. E-mail: wuhao1@sslab.org.cn, zhixiong_li@csu.edu.cn

**We investigate spin-Hall-like magnon edge transport in a synthetic antiferromagnetic skyrmion lattice composed of two antiferromagnetically coupled skyrmion lattice layers with opposite magnetic textures. Based on a relaxed bilayer texture from micromagnetic simulations, we construct the bosonic Bogoliubov-de Gennes Hamiltonian within linear spin-wave theory and calculate the bulk and strip magnon spectrum. We find counterpropagating in-gap edge modes with opposite layer polarization, whose layer-resolved propagation is further confirmed by dynamical micromagnetic simulations. A symmetry analysis shows that the fully coupled system lacks the pseudo-time-reversal symmetry required for a genuine bosonic $Z_2$ topological phase. Thus, the observed edge modes are not $Z_2$-protected helical magnon edge states, but layer-polarized, spin-Hall-like modes originating from the opposite Hall tendencies of the two skyrmion lattice layers. These results establish synthetic antiferromagnetic skyrmion lattices as a platform for spin-Hall-like magnon transport beyond a strict bosonic $Z_2$ classification.**

# I. INTRODUCTION

Topological magnonics has become an important branch of condensed-matter physics, extending the concepts of band topology and protected edge transport from fermionic systems to bosonic collective excitations. In magnetic systems, magnons acquire nontrivial Berry curvature through spin-orbit coupling, noncollinear magnetic order, or real-space magnetic textures, generating topological magnon bands, chiral edge states, and transverse thermal transport [1-11].

Among available platforms, skyrmion-based magnetic textures are particularly attractive, as they naturally generate emergent fields from the magnetic texture [12,13]. In a conventional **skyrmion lattice (SkL)**, the nonzero topological charge produces a finite net emergent magnetic field for magnons, leading to chiral magnon edge states and a thermal Hall response [12-18]. In our recent work on **skyrmionium lattice (SkML)**, we showed that topological magnon bands and chiral edge states can persist even when the net topological charge vanishes [19]. These results highlight skyrmion-based magnetic textures as a versatile route to Hall-type topological magnon transport.

A natural next question is whether a bilayer SkL texture can realize a magnonic analogue of spin-Hall transport. In a **synthetic antiferromagnetic skyrmion lattice (SAF-SkL)**, two SkL layers with opposite magnetization textures are coupled antiferromagnetically. Since the two layers carry opposite texture-induced transverse responses, one may expect counterpropagating magnon channels associated with opposite layer polarization. The layer degree of freedom can then play the role of an effective pseudospin, suggesting a spin-Hall-like magnon response with a largely compensated total Hall signal but a finite layer-polarized transverse response. However, this question cannot be answered from real-space intuition alone. In bosonic **Bogoliubov-de Gennes (BdG)** systems, the topological classification differs fundamentally from the fermionic tenfold-way paradigm [20-28]. In particular, a genuine bosonic $Z_2$ topological phase requires a pseudo-time-reversal symmetry whose square is $-1$ [27,28], rather than the ordinary bosonic time-reversal symmetry with square $+1$. Therefore, for the SAF-SkL, the central issue is not merely whether the two layers carry opposite Hall tendencies, but whether the full coupled bilayer magnon Hamiltonian satisfies the symmetry requirements needed for a genuine bosonic $Z_2$ characterization.

In this work, we address these questions by combining micromagnetic simulations, linear spin-wave theory, strip-spectrum calculations, and dynamical wave-packet simulations. Starting from a relaxed SAF-SkL texture, we derive the bosonic BdG Hamiltonian and obtain the bulk magnon spectrum. We find low-energy collective modes with linear dispersion near the Γ point, reflecting the antiferromagnetically coupled bilayer dynamics. In strip geometry, we identify counterpropagating in-gap edge modes with opposite layer polarization, and confirm their propagation by micromagnetic simulations. At the same time, our symmetry analysis shows that the fully coupled SAF-SkL lacks the pseudo-time-reversal symmetry required for a genuine bosonic $Z_2$ phase. The observed edge modes should therefore be interpreted not as $Z_2$-protected helical magnon edge states, but as layer-polarized, spin-Hall-like edge modes inherited from the opposite Hall tendencies of the two constituent SkL layers. Our results establish the SAF-SkL as a textured-magnetic platform for spin-Hall-like magnon edge transport without relying on a strict bosonic $Z_2$ classification.

## II. MODEL AND BULK SPECTRUM

We consider a SAF-SkL with Néel-type textures realized on a two-dimensional square spin lattice, where the skyrmions form a triangular crystal. As illustrated in the left panel of Fig. 1(a), the system consists of two antiferromagnetically coupled SkL layers. The right panel presents an enlarged view of a single magnetic unit cell, highlighting the detailed bilayer spin configuration. The system is described by the spin Hamiltonian

$$H = -\sum_{<i,j>} J_{ex} \vec{S}_i \cdot \vec{S}_j - \sum_{<i,i'>} J_{af} \vec{S}_i \cdot \vec{S}_{i'} + \sum_{<i,j>} D(\hat{z} \times \vec{r}_{ij}) \cdot (\vec{S}_i \times \vec{S}_j) - K \sum_i (S_i^z)^2, \quad (1)$$

where $\vec{S}_i$ is the spin-$S$ operator at any site $i$. Here, $<i,j>$ runs over nearest-neighbor sites within each layer, whereas $<i,i'>$ denotes vertically aligned sites in the two different layers. The parameter $J_{ex}$ represents the intralayer nearest-neighbor exchange interaction; $J_{af}$ characterizes the interlayer antiferromagnetic coupling; $D$ is the strength of the interfacial Néel-type Dzyaloshinskii-Moriya interaction; and $K$ denotes the perpendicular uniaxial anisotropy.

Unless otherwise specified, the micromagnetic simulations are performed with the following material parameters: the saturation magnetization $M_s = 0.58\times10^6$ A/m, the intralayer exchange stiffness $A_{ex} = 1.5\times10^{-11}$ J/m, the interlayer antiferromagnetic exchange stiffness $A_{af} = -1.5\times10^{-12}$ J/m, the interfacial DMI constant $D_{ind} = 3$mJ/m$^2$, and the perpendicular magnetic anisotropy $K_u$ = 0.36 × $10^6$ J/m$^3$. The computational cell size is chosen as 2×2×2 nm$^3$ , corresponding to a discretization length $a$ = 2 nm. These micromagnetic parameters are mapped to the Hamiltonian parameters in Eq. (1) as: the spin quantum number $S = Msa^3/g\mu_B$ with $g$-factor $g$=2, the intralayer exchange $J_{ex} = 2A_{ex}a/S^2$, the interlayer exchange $J_{af} = 2A_{af}a/S^2$, the DMI $D=D_{ind}a^2/S^2$, and the anisotropy $K= K_u a^3/S^2$ [29-32].

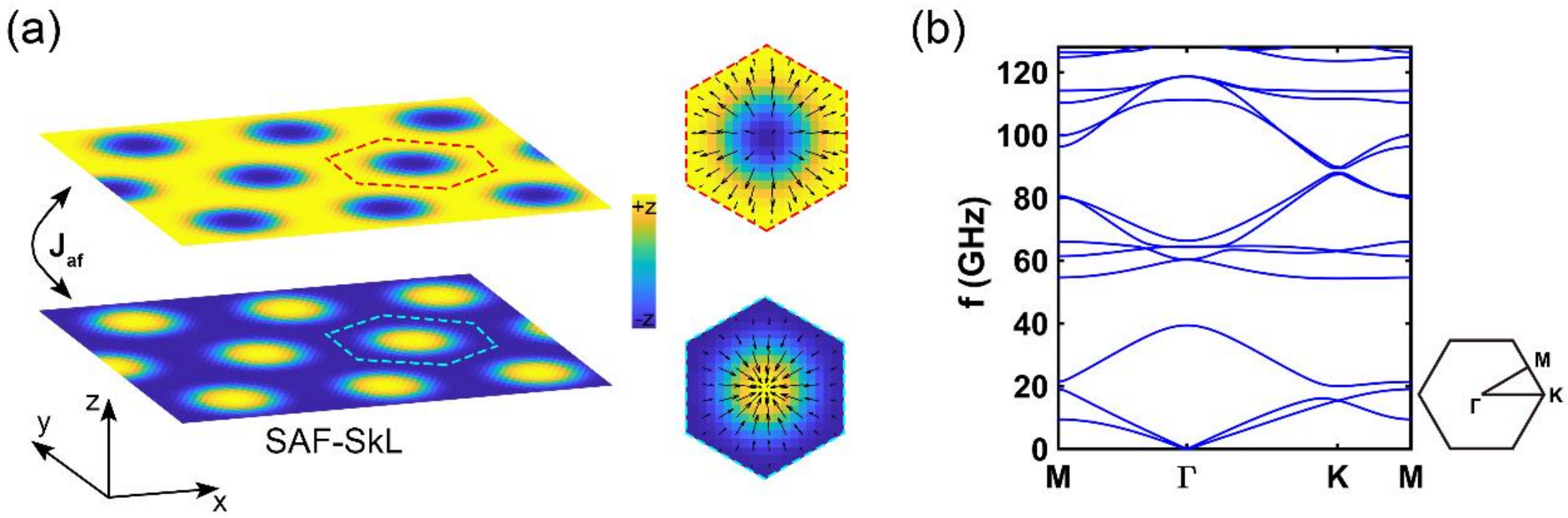

**FIG. 1. Bilayer SAF-SkL texture and its bulk magnon spectrum.** (a) Spin configuration of the SAF-SkL composed of two antiferromagnetically coupled skyrmion-crystal layers. The left panel shows the bilayer texture, where the arrows indicate the interlayer antiferromagnetic coupling $J_{af}$ . The upper and lower layers are color-coded by the out-of-plane spin component. The right panel shows enlarged views of the magnetic unit cell in the two layers, as marked by the dashed hexagons in the left panel. (b) Bulk magnon band structure of the SAF-SkL along the high-symmetry path M-Γ-K-M. Only the lowest fourteen bands are shown, highlighting the low-energy collective excitations of the bilayer skyrmion crystal.

Using these parameters, we stabilize a SAF-SkL in MuMax3, with an equilibrium nearest-neighbor skyrmion spacing of d=40 nm. To maintain consistency with the spin Hamiltonian Eq. (1), the demagnetization field is turned off in MuMax3. After a relaxation, the system converges to the minimum energy SAF-SkL state texture shown in Fig. 1(a). Based on this relaxed magnetic texture, we then derive the magnon excitations within linear spin-wave theory. Specifically, we perform the Holstein-Primakoff boson expansion around the noncollinear SAF-SkL state [1,33,34], and subsequently diagonalize the resulting bosonic BdG Hamiltonian by means of a para-unitary transformation [35-37]. This procedure yields both the bulk magnon dispersion $E(\boldsymbol{k})$ and the corresponding eigenstates $\psi(\boldsymbol{k})$. Technical details are presented in **Appendix A**.

The resulting bulk magnon spectrum is shown in Fig. 1(b), where we plot the lowest fourteen bands along the high-symmetry path M-Γ-K-M (right corner). Since the SAF-SkL consists of two coupled skyrmion lattice layers, its band structure may be viewed as evolving from two copies of the single-layer SkL spectrum. In the decoupled limit, these two copies are trivially degenerate. Once the interlayer antiferromagnetic coupling is introduced, however, the degeneracy is lifted and the two layers hybridize into genuine bilayer collective modes characteristic of the SAF-SkL. A prominent feature of the low-energy spectrum is the appearance of linearly dispersing branches in the vicinity of the Γ point, which is in contrast to the quadratic, parabolic dispersion of the single-layer SkL near the Γ point [14]. These modes are naturally interpreted as collective excitations associated with the antiferromagnetically coupled bilayer dynamics, and therefore constitute a characteristic signature distinguishing the SAF-SkL from an isolated single-layer SkL.

## III. EDGE STATES AND LAYER POLARIZATION

For a SkL, the nonzero topological charge generates a nonzero net emergent magnetic field for magnons. As a result, propagating magnons experience a transverse deflection analogous to a Lorentz force, giving rise to chiral topological edge states characterized by nonzero Chern numbers [12-18]. From this perspective, a synthetic antiferromagnetic skyrmion lattice (SAF-SkL), formed by antiferromagnetically coupling two oppositely polarized SkL layers, may appear to be a natural bosonic analogue of a spin-Hall topological phase: each constituent SkL layer tends to support a magnon Hall response of opposite sign, so that magnons in the two layers are deflected in opposite transverse directions. To study the edge excitations and reveal the Hall response of the SAF-SkL, we consider a strip geometry that is open along the y direction and periodic along the $x$ direction. The resulting strip magnon spectrum is shown in Fig. 2(a). The color of each branch encodes the normalized expectation value of the magnon transverse coordinate, $\langle y\rangle/L_y$, which directly characterizes the spatial localization of the corresponding mode. For each magnon eigenstate $\psi(\boldsymbol{k}) = (u(\boldsymbol{k}), v(\boldsymbol{k}))^T$, we first evaluate its real-space magnon density at site $i$ by $\rho_i \propto |u_i|^2 + |v_i|^2$, and then compute the average transverse position from this distribution. As shown in Fig. 2(a) , in-gap branches localized near the upper or lower edge are clearly identified as edge modes. These edge states appear as nearly doubly degenerate counterpropagating branches, predominantly associated with the two constituent layers. To make the two opposite branches visually distinguishable, we apply a very small out-of-plane magnetic field $h_z \approx 10mT$ in the strip calculation, which slightly lifts their exact overlap without changing the qualitative structure of the spectrum. The inset enlarges the dotted-circled region in Fig. 2(a), making this small splitting more evident.

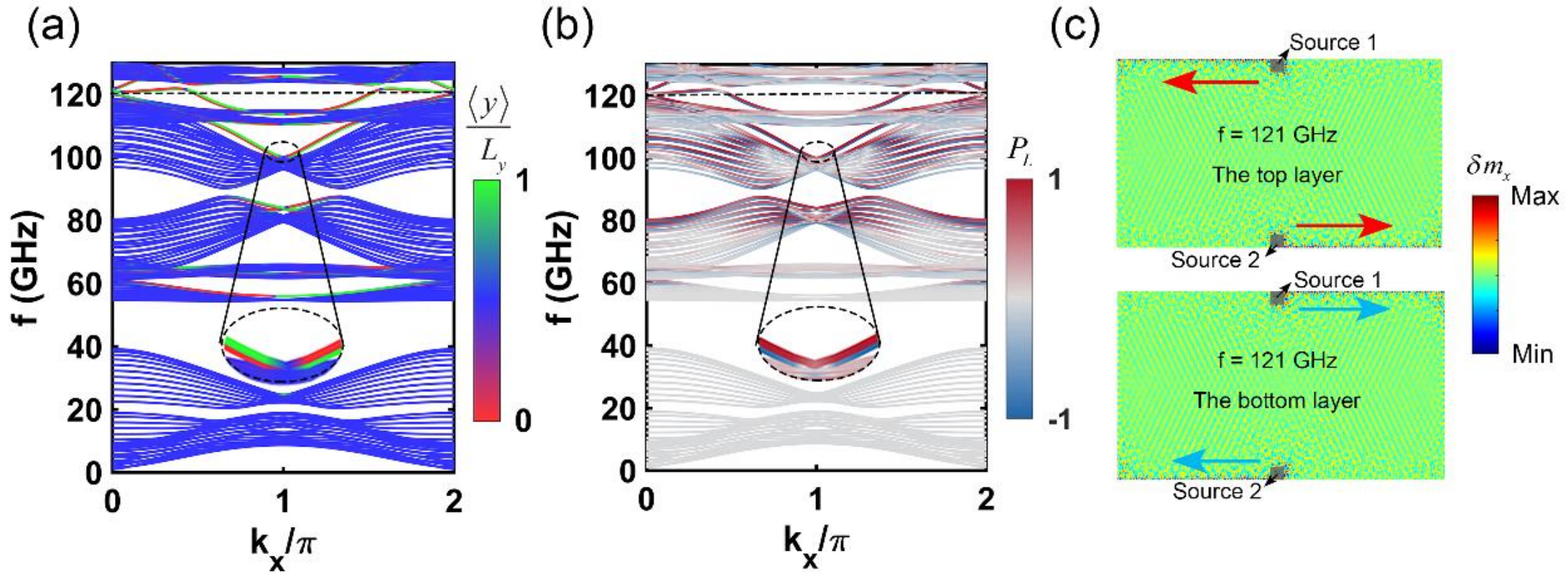

FIG. 2. Edge states and layer polarization in the SAF-SkL. (a) Strip magnon spectrum colored by the normalized transverse position $\langle y \rangle / L_y$ . The dashed horizontal lines indicate the excitation frequency used in the micromagnetic simulation. (b) The same spectrum colored by the layer polarization $P_L$ . (c) Micromagnetic snapshots of $\delta m_x(\boldsymbol{r}, t)$ at 121 GHz, showing counterpropagating edge transport in the two layers.

For the $n$-th bosonic BdG eigenstate $\psi(\boldsymbol{k}) = (u(\boldsymbol{k}), v(\boldsymbol{k}))^T$ , we define the layer polarization through the layer-label matrix $\Lambda_L = \mathrm{diag}(\eta_1, \dots, \eta_N)$, where $\eta_i = +1\ (-1)$ for sites on the top (bottom) layer. The layer polarization of the $n$-th magnon band is then defined as

$$P_L(n,\mathbf{k}) = \frac{u^\dagger(\mathbf{k})\Lambda_L u(\mathbf{k}) + v^\dagger(\mathbf{k})\Lambda_L v(\mathbf{k})}{u^\dagger(\mathbf{k})u(\mathbf{k}) + v^\dagger(\mathbf{k})v(\mathbf{k})} . \tag{2}$$

Here $P_L$ characterizes the layer-resolved spectral weight of a magnon eigenmode and should not be confused with the circular polarization or spin angular momentum of magnons. In the present SAF-SkL, the layer degree of freedom acts as an effective pseudospin. This quantity ranges from +1 to −1, corresponding to modes fully localized on the top and bottom layers, respectively. In Fig. 2(b), we plot the same strip spectrum but now color the branches according to $P_L$. One clearly sees that the nearly overlapping in-gap edge branches carry opposite layer polarizations and are therefore predominantly localized in different layers. Owing to the small symmetry-breaking field $h_z \approx 10mT$, the two branches are no longer exactly degenerate: the branch with $P_L = +1$ is shifted slightly upward in frequency relative to the branch with $P_L = -1$.

This tiny splitting is highlighted in the inset of Fig. 2(b). The spectrum therefore shows explicitly that the edge modes of the SAF-SkL are not only edge-localized but also strongly layer polarized.

To provide a dynamical real-space verification of these layer-polarized edge states, we further perform micromagnetic simulations using MuMax3 [29]. Two sinusoidal microwave sources are applied at the two $y$-direction edges, and each source excites both magnetic layers simultaneously. The excitation frequency is chosen to be 121 GHz, which lies inside the edge-mode window highlighted by the dashed lines in Figs. 2(a) and 2(b). We then let the system evolve in time and record snapshots of the dynamic $x$-component of the magnetization, $\delta m_x(\boldsymbol{r}, t)$, during magnon propagation, as shown in Fig. 2(c). The simulated dynamics clearly demonstrate a spin-Hall-like propagation pattern: edge excitations in the top and bottom layers propagate with opposite chirality. In other words, the two layers support counterpropagating edge transport channels with opposite layer polarization, in full agreement with the strip spectrum analysis.

These results show that the SAF-SkL supports observable layer-polarized edge excitations. In this sense, the SAF-SkL realizes a magnon spin-Hall-like state, in which the layer degree of freedom plays the role of an effective pseudospin.

## IV. SYMMETRY ANALYSIS AND ABSENCE OF A BOSONIC $Z_2$ INVARIANT

However, whether such a bilayer system realizes a magnonic spin Hall Effect and is genuinely characterized by a $Z_2$ invariant cannot be inferred from this intuitive picture alone; it must instead be determined from the symmetry class of the corresponding bosonic BdG Hamiltonian.

Important progress has been made in the topological classification of topological magnons [26-28]. A key conclusion is that, in bosonic systems, the tenfold-way classification formulated for fermions, based on time-reversal symmetry, chiral symmetry, and particle-hole symmetry, collapses into a threefold way, in which only time-reversal symmetry remains fundamental [28]. For pseudo-Hermitian magnon BdG Hamiltonians, the classification is therefore determined solely by whether time-reversal symmetry is present and, if so, whether the square of the time-reversal operator is +1 or −1. Accordingly, pseudo-Hermitian magnon BdG Hamiltonians fall into three relevant classes: $C_0$, corresponding to the absence of time-reversal symmetry; $R_0$, corresponding to a bosonic time-reversal symmetry with $\Theta^2$=+1; and $R_4$, corresponding to a pseudo-time-reversal symmetry with $(\Theta')^2$=−1. This topological classification is presented in Table I, as summarized from Ref. [28]. In the two-dimensional (d=2) system, these three classes support, respectively, a Z invariant, no symmetry-protected topological invariant, and a $Z_2$ invariant. Therefore, to determine the topological classification of the SAF-SkL, it is sufficient to identify the time-reversal symmetry realized by its magnon Hamiltonian. Note that the time-reversal symmetry with $\Theta^2$=−1 considered here is a pseudo-time-reversal symmetry, because for bosonic systems the genuine time-reversal operator always satisfies $\Theta^2$=+1.

Table I. Topological classification of pseudo-Hermitian bosonic BdG Hamiltonians relevant to magnon systems. Here Θ=0, +1, and −1 denote, respectively, the absence of time-reversal symmetry, conventional bosonic time-reversal symmetry with $\Theta^2$=+1, and pseudo-time-reversal symmetry with $(\Theta')^2$=−1. The corresponding classifying spaces and topological invariants are listed for spatial dimensions d=0-3.

| $\Theta$ | Classifying space | $d$=0 | $d$=1 | $d$=2 | $d$=3 |
|---|---|---|---|---|---|
| 0 | $C_0$ | Z | - | Z | - |
| 1 | $R_0$ | Z | - | - | - |
| -1 | $R_4$ | 2Z | - | $Z_2$ | $Z_2$ |

For an ideal bilayer magnetic system composed of two oppositely polarized textures coupled antiferromagnetically, the bosonic BdG Hamiltonian can generally be written in the form [27],

$$H_{BdG}(\boldsymbol{k}) = \begin{pmatrix} h_1(\boldsymbol{k}) & h_2(\boldsymbol{k}) & \Delta_2(\boldsymbol{k}) & \Delta_1(\boldsymbol{k}) \\ h_2^\dagger(\boldsymbol{k}) & h_1^*(-\boldsymbol{k}) & \Delta_1^*(-\boldsymbol{k}) & \Delta_2^\dagger(\boldsymbol{k}) \\ \Delta_2^\dagger(\boldsymbol{k}) & \Delta_1^*(-\boldsymbol{k}) & h_1^*(-\boldsymbol{k}) & h_2^\dagger(-\boldsymbol{k}) \\ \Delta_1(\boldsymbol{k}) & \Delta_2(\boldsymbol{k}) & h_2(-\boldsymbol{k}) & h_1(\boldsymbol{k}) \end{pmatrix}, \tag{3}$$

where $h_{1,2}(k)$ and $\Delta_{1,2}(k)$ are block matrices in the internal sublattice space. This Hamiltonian respects the bosonic time-reversal symmetry

$$\Sigma_z H_{BdG}(-\boldsymbol{k})\Theta - \Theta\Sigma_z H_{BdG}(\boldsymbol{k}) = 0, \tag{4}$$

where $\Sigma_z = \sigma_z \otimes 1_N$, and the bosonic time-reversal takes the form

$$\Theta = (\sigma_z \otimes \sigma_x \otimes 1_{N/2})K. \tag{5}$$

Here $\sigma_z$ acts on the particle-hole space, $\sigma_x$ acts on the layer degree of freedom, and $K$ denotes complex conjugation. Physically, Θ combines the interchange of the two layers with a $C_2$ rotation of the spins. In the special case $\Delta_2(k)=0$, the Hamiltonian further possesses a pseudo-time-reversal symmetry [23, 27],

$$\Theta' = (\sigma_z \otimes i\sigma_y \otimes 1_{N/2})K, \tag{6}$$

with $(\Theta')^2=-1$. Physically, this operation interchanges the spin degrees of freedom in the upper and lower layers, introduces an additional minus sign, and subsequently applies a $C_2$ rotation to each spin. It guarantees bosonic Kramers pairs at the time-reversal-invariant momenta and thereby allows for a genuine $Z_2$ characterization.

The SAF-SkL, however, does not satisfy this special condition. Because the skyrmion texture is intrinsically noncollinear and the local spin directions are not globally locked to a single easy axis, the anomalous interlayer block $\Delta_2(\mathbf{k})$ is generically finite. Consequently, the pseudo-time-reversal symmetry $(\Theta')^2=-1$ is absent, even though the conventional bosonic time-reversal symmetry $\Theta^2=+1$ remains. According to the threefold classification of bosonic pseudo-Hermitian BdG Hamiltonians, the two-dimensional SAF-SkL therefore belongs to class $R_0$. This immediately implies that the fully coupled SAF-SkL is not a bosonic analogue of a class-AII topological insulator and cannot be characterized by a symmetry-protected $Z_2$ invariant. At the same time, the ordinary band Chern number is not the appropriate topological descriptor for the full bilayer system, since the two layers contribute oppositely and the total Hall response cancels. Therefore, under interlayer coupling, neither the conventional Chern number nor the bosonic $Z_2$ index remains an appropriate strict topological classifier.

The absence of a symmetry-protected $Z_2$ invariant does not imply the absence of edge transport. Rather, it means that the counterpropagating in-gap branches in Fig. 2 are not protected by the bosonic $Z_2$ mechanism associated with$(\Theta')^2=-1$. In the SAF-SkL, the two skyrmion layers carry opposite magnetic textures and thus tend to generate opposite Berry-curvature-induced transverse responses. Even though finite interlayer coupling hybridizes the two layers and removes exact layer conservation, the corresponding edge modes can still retain strong layer polarization. Therefore, the edge branches observed in Fig. 2 should be interpreted as layer-polarized, spin-Hall-like edge modes rather than as $Z_2$-protected helical magnon edge states. The fully coupled SAF-SkL is thus not a genuine bosonic analogue of a class-AII topological insulator, but it nevertheless provides a platform for spin-Hall-like magnon edge transport with opposite layer polarization.

## V. ORIGIN OF THE EDGE MODES IN THE SAF-SkL

The results above show that the SAF-SkL supports well-defined layer-polarized in-gap edge modes. Since the SAF-SkL does not possess a strict Z or $Z_2$ topological classification, these edge modes are not protected in the same sense as Chern edge states or $Z_2$-protected helical edge states. This raises an important question: where do the edge modes come from? As shown below, these edge states can be traced back to the topological edge states in SAF-SkL in the absence of interlayer coupling, or equivalently, in the decoupled single-layer limit. To clarify this point, we examine how the bulk and strip spectrum evolve as the interlayer antiferromagnetic coupling is gradually increased from the decoupled limit.

We calculate the bulk and strip magnon spectrum for several values of the interlayer exchange stiffness as shown in Fig. 3(a)-3(d), corresponding to $A_{af} = 0$, $-0.3\times10^{-12}$, $-0.6\times10^{-12}$, $-0.9\times10^{-12}$ J/m, respectively. In each panel, the upper spectrum is the bulk magnon dispersion colored by the layer polarization $P_L$, while the lower spectrum is the strip magnon dispersion colored by the normalized transverse position $\langle y\rangle/L_y$. A small out-of-plane magnetic field $h_z \approx 10mT$ is included in all calculations to slightly lift the near degeneracy of the layer-polarized branches and make them visually distinguishable.

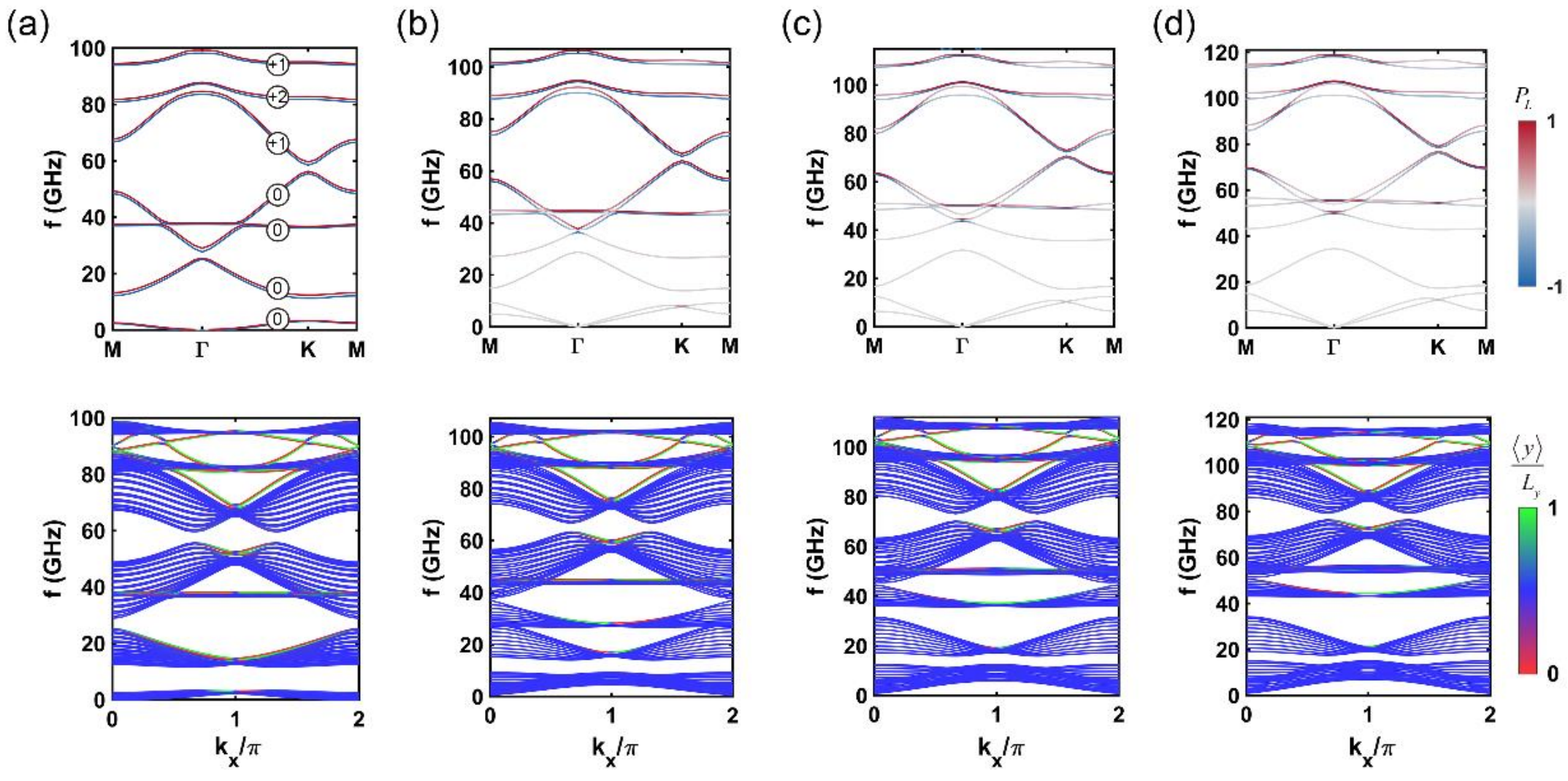

FIG. 3. Evolution of the bulk and strip magnon spectrum with increasing interlayer antiferromagnetic coupling. Panels (a)-(d) correspond to $A_{af}$ = 0, -0.3×10$^{-12}$, -0.6×10$^{-12}$, -0.9×10$^{-12}$ J/m, respectively. In each panel, the upper spectrum shows the bulk magnon bands colored by the layer polarization $P_L$, while the lower spectrum shows the strip magnon spectrum colored by the normalized transverse position $\langle y\rangle/L_y$. A small out-of-plane magnetic field $h_z \approx 10mT$ is introduced in all calculations to slightly lift the nearly degenerate branches. The spin Chern number is labeled for each pair of bands for the decoupled case (a).

We first consider the decoupled limit $A_{af}$=0, where the two skyrmion lattice layers are completely independent. In this limit, each layer can be characterized by its own Chern number. Since the two layers are related by opposite magnetization textures, their magnon spectrum are identical, whereas their Berry curvatures and Chern numbers have opposite signs. Denoting the Chern numbers of the top and bottom layers by $C_t$ and $C_b$, one has $C_t$=−$C_b$. The method of obtaining the Chern number is presented in **Appendix B.** In this special limit, the layer index is regarded as an exactly conserved effective pseudospin $\sigma_z$, thus the natural topological quantity is the layer-resolved spin Chern number [23,38],

$$C_s = (C_t - C_b)/2. \quad (7)$$

In other words, the decoupled SAF-SkL belongs to a spin-Chern phase in the same sense as bilayer systems with conserved pseudospin: it supports counterpropagating, layer-polarized magnon edge modes without a net thermal Hall response. Importantly, this use of the spin Chern number is justified here precisely because the two layers are dynamically independent and the layer pseudospin is conserved. As shown in Fig. 3(a), we compute the bulk and strip magnon spectrum of the SAF-SkL in the decoupled limit, with the upper and lower panels displaying the bulk and strip spectrum, respectively. In the bulk spectrum, the spin Chern number is labeled for each pair of bands, while the layer polarization defined in Eq. (2) is colored. In the strip spectrum, another colormap is used to represent the magnon transverse coordinate $\langle y\rangle/L_y$, clearly revealing edge states within the gap between 80 and 100 GHz. In all these calculations, a small out-of-plane magnetic field $h_z \approx 10mT$ is introduced to slightly lift the original double degeneracy and thus make the bands distinguishable.

We then turn on a finite interlayer antiferromagnetic coupling. As shown in Figs. 3(b)-3(d), increasing $A_{af}$ shifts the magnon frequencies and gradually hybridizes the two layer-resolved spectrum. Consequently, the layer degree of freedom is no longer an exactly conserved quantum number. The spin Chern number in Eq. (7), which relies on exact layer conservation, therefore ceases to be a strict topological invariant of the fully coupled SAF-SkL. This is consistent with the symmetry analysis in Sec. IV: the SAF-SkL belongs to class R0 and is not protected by either a conventional Chern number of the full bilayer system or a bosonic $Z_2$ invariant. Nevertheless, the edge modes do not disappear immediately once the interlayer coupling is introduced. Instead, the strip spectrum in Figs. 3(b)-3(d) show that the in-gap edge branches evolve continuously from those in the decoupled limit and remain clearly visible over the coupling range considered here. These modes should therefore be understood as spectral descendants, or remnants, of the topological edge modes of the two isolated SkL layers. Their existence at finite coupling is no longer guaranteed by a strict bulk topological invariant, but their layer-polarized character can survive as long as the interlayer hybridization does not completely erase the layer texture inherited from the decoupled limit.

This interpretation reconciles the strip spectrum results with the symmetry classification. The fully coupled SAF-SkL is not a genuine bosonic $Z_2$ topological insulator, and its edge modes should not be described as $Z_2$-protected helical magnon edge states. Rather, they are layer-polarized, spin-Hall-like edge modes originating from the opposite Hall tendencies of the two constituent SkL layers. The finite interlayer coupling removes strict topological protection, but it does not immediately destroy the edge transport channels. As a result, the SAF-SkL can exhibit a non-quantized spin-Hall-like magnon response, characterized by counterpropagating edge transport with opposite layer polarization, even in the absence of a strict Z or $Z_2$ topological classification.

## VI. HALL-LIKE TRANSPORT PICTURE OF SKYRMION-BASED MAGNON SYSTEMS

Previous studies have established topological magnon transport in conventional SkLs, where the nonzero net emergent flux associated with the magnetic texture gives rise to a transverse magnon response and chiral edge states [12-18]. In recent work on the SkML, it further showed that topological magnon states can persist even when the net topological charge vanishes [19]. In the present SAF-SkL system, we find counterpropagating edge modes with opposite layer polarization. Taken together, these results suggest that different skyrmion-based magnetic superstructures can support distinct Hall-like magnon transport behaviors. This comparison should be understood only as a physical analogy at the level of transverse response, rather than as a strict identification of microscopic mechanisms or symmetry classes. The common element is that, in all these systems, the magnetic texture or band geometry generates an effective transverse deflection of magnon wave packets, which can be described within the Berry-curvature framework.

For a magnon wave packet in the $n$-th band, the semiclassical equation of motion takes the form [3,4,39,40]

$$\dot{\boldsymbol{r}} = \frac{1}{\hbar}\nabla_k \varepsilon_n(\boldsymbol{k}) - \dot{\boldsymbol{k}} \times \boldsymbol{\Omega}_n(\boldsymbol{k}), \tag{8}$$

where $\varepsilon_n(\boldsymbol{k})$ is the band energy and $\boldsymbol{\Omega}_n(\boldsymbol{k})$ is the Berry curvature. The second term gives rise to an anomalous transverse velocity whenever $\dot{\boldsymbol{k}} \neq 0$. As a consequence, both electronic Hall transport and magnon thermal Hall transport can be expressed in Berry-curvature form. Schematically, the magnon thermal Hall conductivity can be written as [3,4,18,41],

$$\kappa_{xy} \propto \sum_n \int \frac{d^2k}{(2\pi)^2} \Omega_n(\boldsymbol{k}) W[n_B(\varepsilon_n)], \tag{9}$$

which is structurally analogous to the electronic Hall conductivity [40,42],

$$\sigma_{xy} \propto \sum_n \int \frac{d^2k}{(2\pi)^2} \Omega_n(\boldsymbol{k}) f(\varepsilon_n). \tag{10}$$

Here, the heat current replaces the charge current, and the Bose distribution replaces the Fermi distribution. Although the microscopic carriers and driving forces are different, these expressions highlight a common geometric origin of transverse transport. From this viewpoint, the Hall-type responses in skyrmion-based lattices and those in electronic systems share the same Berry-curvature framework, even though their microscopic origins are different.

In Fig. 4, we provide an intuitive schematic illustration of this picture between the three skyrmion-based lattices (SkL, SkML, SAF-SkL) and and three Hall-type responses: the ordinary Hall effect (HE), anomalous Hall effect (AHE), and spin Hall effect (SHE). Below, we clarify each case with explicit physical reasoning and, where appropriate, with equation-level analogies.

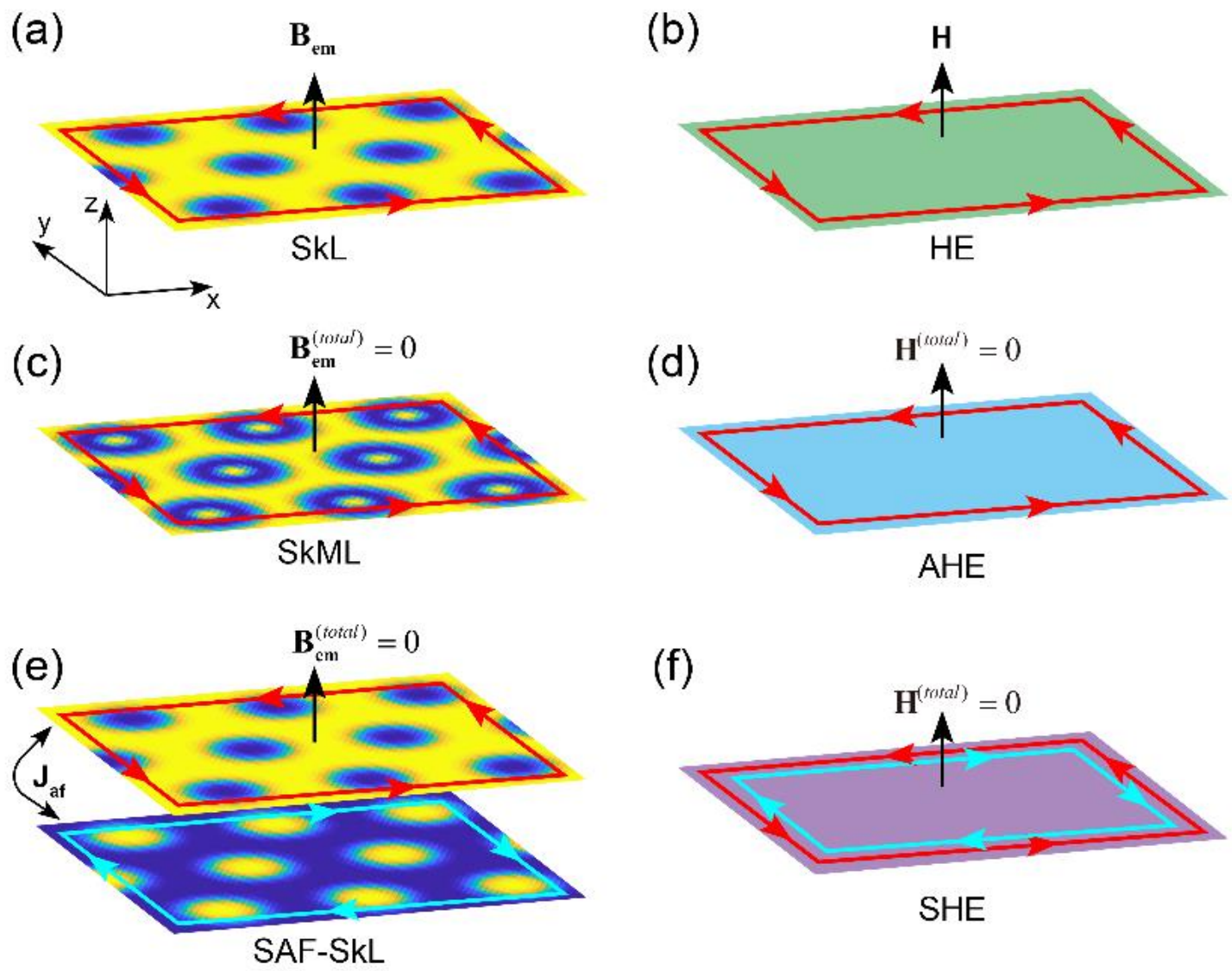

FIG. 4. **Schematic analogy between skyrmion-based lattices and Hall-type responses. (a), (b)** Conventional SkL and the ordinary Hall effect, both characterized by a finite net effective flux and a corresponding transverse response. **(c), (d)** SkML and the anomalous Hall effect, where a transverse response survives despite vanishing net flux owing to Berry curvature and band topology. **(e), (f)** SAF-SkL and a spin-Hall-like response, where opposite layer-resolved transverse currents largely cancel in the total channel but remain finite in the layer-polarized channel.

For a conventional SkL, the nonzero topological charge density generates a finite emergent magnetic field, which acts on magnons as an effective Lorentz-like field. Magnon wave packets are therefore deflected transversely [Fig. 4(a)], giving rise to a finite thermal Hall response [12-18]. In this sense, the magnon Hall effect in a SkL is the natural magnonic analogue of the ordinary Hall effect in electronic systems, where a nonzero net magnetic flux produces a transverse current [Fig. 4(b)]. The key feature in both cases is the presence of a finite net effective flux, which leads to Berry curvature of a definite overall sign and hence to a nonvanishing transverse response.

The situation is qualitatively different in the SkML. Because a skyrmionium carries zero net topological charge, the total emergent field averaged over a magnetic unit cell vanishes. Nevertheless, as shown in our previous study [19], the magnon bands of the SkML can still carry nonzero Chern numbers and support chiral edge modes. The resulting transverse response survives even in the absence of a net emergent flux [Fig. 4(c)]. This behavior is closely analogous to the anomalous Hall effect, in which a Hall response arises from band topology and Berry curvature even when the net external magnetic field is zero [Fig. 4(d)]. From this perspective, the SkML may be viewed as a magnetic-texture realization of an anomalous-Hall-like magnon response.

The SAF-SkL provides the corresponding bilayer analogue of a spin-Hall-type response. It is constructed from two antiferromagnetically coupled SkL layers with opposite magnetization textures and opposite topological charges. As a result, each layer individually tends to support a transverse magnon response of opposite sign. Magnons in the top layer are deflected in one transverse direction, whereas magnons in the bottom layer are deflected in the opposite direction. Consequently, the total transverse thermal current from the two layers can largely cancel,

$$j_y^{(\mathrm{tot})} \propto j_y^{(\mathrm{top})} + j_y^{(\mathrm{bot})} \approx 0, \tag{11}$$

while a layer-polarized ("spin-like") transverse current remains finite,

$$j_y^{\mathrm{sp}} \propto j_y^{(\mathrm{top})} - j_y^{(\mathrm{bot})} \neq 0. \tag{12}$$

This is directly analogous to the structure of the electronic spin Hall effect [43,44], in which opposite spin sectors carry opposite transverse currents so that the net charge Hall current vanishes while the spin current survives. In the SAF-SkL, the layer degree of freedom plays the role of an effective pseudospin.

We emphasize again that this interpretation refers to the structure of the transverse response, not to a strict symmetry-protected topological classification. As shown in Sec. IV, the fully coupled SAF-SkL does not possess the pseudo-time-reversal symmetry required for a genuine bosonic $Z_2$ phase. Therefore, the terminology "spin-Hall-like" should be understood as a physically transparent description of the observed layer-polarized counterpropagating edge transport, rather than as implying a bosonic analogue of a class-AII topological insulator in the strict sense.

Overall, SkL, SkML, and SAF-SkL provide three representative examples of how nontrivial magnetic textures can generate qualitatively different Hall-like magnon transport behaviors. This viewpoint offers a compact physical framework for organizing the transverse magnon responses induced by skyrmion-based magnetic superstructures.

## VII. CONCLUSION AND DISCUSSION

In summary, we have shown that a SAF-SkL supports layer-polarized, spin-Hall-like magnon edge transport. Although the fully coupled SAF-SkL lacks the pseudo-time-reversal symmetry required for a genuine bosonic $Z_2$ phase, it still supports counterpropagating in-gap edge modes with opposite layer polarization. This result is particularly relevant for magnonic applications because synthetic antiferromagnetic textures combine compensated magnetization with layer-resolved dynamics. The nearly compensated magnetic structure can suppress stray fields, while the opposite skyrmion textures in the two layers provide opposite transverse magnon responses. As a result, the total Hall-like signal may be largely compensated, whereas a layer-polarized magnon current remains finite. The layer degree of freedom therefore provides an effective pseudospin channel for spin-Hall-like magnon transport in engineered multilayer magnets.

More broadly, our results support a unified picture of Hall effects induced by topological magnetic textures: the SkL, SkML, and SAF-SkL may be associated, respectively, with ordinary-Hall-like, anomalous-Hall-like, and spin-Hall-like magnon responses. Future work may explore whether nonlinear spin-wave dynamics, or interaction effects in textured magnetic multilayers can lead to correlated magnonic states beyond the linear spin-wave regime.

## ACKNOWLEDGMENTS

We thank Peng Yan for discussion. This work was supported by the National Key Research and Development Program of China (Grant Nos. 2022YFA1402801), the National Natural Science Foundation of China (NSFC, Grant Nos. 52271239, 12204089 and T2495212), the Guangdong Basic and Applied Basic Research Foundation (Grant Nos. 2026B0303000004 and 2024A1515110196), Guangdong Provincial Quantum Science Strategic Initiative (Grant Nos. GDZX2501005 and GDZX2502004), the Natural Science Foundation of Hunan Province of China (Grant No. 2024JJ6113).

## APPENDIX A: LINEAR SPIN-WAVE THEORY FOR THE SAF-SkL

In this Appendix, we present the details of the linear spin-wave calculation used to obtain the bulk magnon spectrum $E_n(\boldsymbol{k})$ and the corresponding bosonic eigenstates $\psi_n(\boldsymbol{k})$ of SAF-SkL. Our derivation follows the standard Holstein-Primakoff formalism for noncollinear magnetic textures and the para-unitary diagonalization of bosonic BdG Hamiltonians [1,33-37].

Based on the magnetic texture and the spin Hamiltonian Eq. (1) of SAF-SkL, for each site $i$ in the magnetic unit cell, we denote the classical spin direction by $\boldsymbol{n}_i = (\sin\theta_i \cos\phi_i, \sin\theta_i \sin\phi_i, \cos\theta_i)$, and introduce a local orthonormal basis $\{\boldsymbol{e}_i^1, \boldsymbol{e}_i^2, \boldsymbol{e}_i^3\}$ with $\boldsymbol{e}_i^3 = \boldsymbol{n}_i$. A convenient choice is $\boldsymbol{e}_i^1 = (\cos\theta_i \cos\phi_i, \cos\theta_i \sin\phi_i, -\sin\theta_i)$, $\boldsymbol{e}_i^2 = (-\sin\phi_i, \cos\phi_i, 0)$, so that $\boldsymbol{e}_i^1 \times \boldsymbol{e}_i^2 = \boldsymbol{e}_i^3$. The spin operator at site i is then written in the local frame as $\boldsymbol{S}_i = \boldsymbol{e}_i^1 \tilde{S}_i^x + \boldsymbol{e}_i^2 \tilde{S}_i^y + \boldsymbol{e}_i^3 \tilde{S}_i^z = \frac{1}{2}\boldsymbol{e}_i^+ \tilde{S}_i^- + \frac{1}{2}\boldsymbol{e}_i^- \tilde{S}_i^+ + \boldsymbol{e}_i^3 \tilde{S}_i^z$, where $\boldsymbol{e}_i^\pm = \boldsymbol{e}_i^1 \pm i\boldsymbol{e}_i^2$ and $\tilde{S}_i^\pm = \tilde{S}_i^x \pm i\tilde{S}_i^y$. We next perform the Holstein-Primakoff transformation in the local frame [1,33,34],

$$\tilde{S}_i^+ = \sqrt{2S - a_i^\dagger a_i}\, a_i, \quad \tilde{S}_i^- = a_i^\dagger \sqrt{2S - a_i^\dagger a_i}, \quad \tilde{S}_i^z = S - a_i^\dagger a_i, \tag{A1}$$

where $a_i^\dagger$ and $a_i$ are bosonic creation and annihilation operators. Keeping only terms up to quadratic order in the bosons, the Hamiltonian becomes

$$H = E_0 + \frac{1}{2}\sum_{ij} A_{ij} a_i^\dagger a_j + B_{ij} a_i^\dagger a_j^\dagger + B_{ij}^* a_i a_j + A_{ij}^* a_i a_j^\dagger, \tag{A2}$$

where $E_0$ is the ground-state energy constant; the summation indices $i, j$ run over all lattice sites in the bilayer magnetic unit cell, including both layers. The coefficients $A_{ij}$ and $B_{ij}$ denote the coupling constants. At this stage, a para-unitary transformation is required to diagonalize the Hamiltonian [35-37]. To facilitate this, we transform the Hamiltonian into momentum space, exploiting the periodicity of the system to simplify the calculations.

We label each site as $i \equiv (\mathbf{R}, \mu)$, where $\mathbf{R}$ is a magnetic Bravais-lattice vector and $\mu = 1, \ldots, N$ labels the $N$ spins inside one bilayer magnetic unit cell. We then introduce the Fourier transform $a_{\boldsymbol{R},\mu} = \sum_{\boldsymbol{k}} e^{i\boldsymbol{k}\cdot(\boldsymbol{R}+\boldsymbol{r}_\mu)} a_{\boldsymbol{k},\mu}$, with $\mathbf{r}_\mu$ the position of the $\mu$-th spin inside the unit cell. In momentum space, after some algebra, the Hamiltonian can be written in a compact matrix form as

$$H = E_0 + \frac{1}{2}\sum_{\boldsymbol{k}} \psi_{\boldsymbol{k}}^\dagger H_{\mathrm{BdG}}(\boldsymbol{k}) \psi_{\boldsymbol{k}}, \tag{A3}$$

where $\psi_{\boldsymbol{k}} = \left(a_{\boldsymbol{k},1}, \dots, a_{\boldsymbol{k},N}, a^\dagger_{-\boldsymbol{k},1}, \dots, a^\dagger_{-\boldsymbol{k},N}\right)^T$ is a 2N-component Nambu spinor, and

$$H_{BdG}(\boldsymbol{k}) = \begin{pmatrix} A(\boldsymbol{k}) & B(\boldsymbol{k}) \\ B^\dagger(\boldsymbol{k}) & A^T(-\boldsymbol{k}) \end{pmatrix}. \tag{A4}$$

The magnon bands are then obtained by the para-unitary diagonalization of the bosonic BdG Hamiltonian. Defining $\sum_z = \sigma_z \otimes I_N$, where $\sigma_z$ is the third Pauli matrix and $I_N$ is the $N \times N$ unit matrix, we solve the generalized eigenvalue problem

$$\Sigma_z H_{\mathrm{BdG}}(\boldsymbol{k})\psi_n(\boldsymbol{k}) = E_n(\boldsymbol{k})\psi_n(\boldsymbol{k}). \tag{A5}$$

The physical magnon bands correspond to the $N$ positive eigenvalues $E_n(\mathbf{k})>0$. The associated eigenvectors are the $2N$-component bosonic wave functions $\psi_n(\boldsymbol{k}) = (u_n(\boldsymbol{k}), v_n(\boldsymbol{k}))^T$, which are normalized by the bosonic metric, $\psi_m^\dagger(\boldsymbol{k})\Sigma_z\psi_n(\boldsymbol{k}) = \delta_{mn}$. Collecting the eigenvectors into the columns of a matrix $T_{\mathrm{k}}$ , one obtains a para-unitary transformation satisfying $T_{\boldsymbol{k}}^\dagger \Sigma_z T_{\boldsymbol{k}} = \Sigma_z$, $T_{\boldsymbol{k}}^{-1} = \Sigma_z T_{\boldsymbol{k}}^\dagger \Sigma_z$, and the quadratic Hamiltonian is diagonalized as

$$H = E_0 + \sum_{\boldsymbol{k},n} E_n(\boldsymbol{k}) \left(\alpha_{\boldsymbol{k},n}^\dagger \alpha_{\boldsymbol{k},n} + \frac{1}{2}\right). \tag{A6}$$

where $\alpha_{\boldsymbol{k},n}^\dagger$ and $\alpha_{\boldsymbol{k},n}$ are the Bogoliubov magnon creation and annihilation operators.

Therefore, the bulk magnon spectrum $E_n(\boldsymbol{k})$ and the corresponding eigenstates $\psi_n(\boldsymbol{k})$ used throughout the main text are obtained directly from the positive energy solutions of Eq. (A5). These bosonic eigenvectors are further used to compute the layer polarization, the strip spectrum edge state localization, and the Berry-curvature-related quantities discussed in the main text.

## APPENDIX B: METHOD FOR OBTAINING THE CHERN NUMBER

Based on the Holstein-Primakoff expansion and para-unitary diagonalization described in Appendix A, we can obtain the eigenvector $\psi_n(\boldsymbol{k})$ of the $n$-th energy band. Then the Chern number for this band can be obtained by [14,19],

$$C_n = \frac{(-1)^{\sigma_n}}{2\pi} \int_{BZ} \Omega_n(\boldsymbol{k}) d^2\boldsymbol{k}, \tag{B1}$$

where $n$ labels the *n-th* magnon band; $\sigma_n = 0(1)$ for positive (negative) energy band; $\Omega_n(\boldsymbol{k})$ is the Berry curvature, given by

$$\Omega_n(k_x, k_y) = \lim_{dk_x, dk_y \to 0} \frac{-i \times log(U_1 U_2 U_3 U_4)}{dk_x dk_y}, \tag{B2}$$

where

$$U_1 = \langle \psi_n(k_x, k_y) | \Sigma_z | \psi_n(k_x + dk_x, k_y) \rangle,$$
$$U_2 = \langle \psi_n(k_x + dk_x, k_y) | \Sigma_z | \psi_n(k_x + dk_x, k_y + dk_y) \rangle,$$
$$U_3 = \langle \psi_n(k_x + dk_x, k_y + dk_y) | \Sigma_z | \psi_n(k_x, k_y + dk_y) \rangle,$$
$$U_4 = \langle \psi_n(k_x, k_y + dk_y) | \Sigma_z | \psi_n(k_x, k_y) \rangle. \quad \text{(B3)}$$

Here, $dk_x(dk_y)$ is the infinitesimal of the $k_x(k_y)$, $\Sigma_z$ is the direct product of the third Pauli matrix and a $N$-dimensional unit matrix.